\documentclass[twocolumn,showpacs,preprintnumbers,amsmath,amssymb,floatfix,prd]{revtex4}
\usepackage{epsfig}
\usepackage{dcolumn}
\begin{document}
\title{Parton-to-Kaon Fragmentation Revisited}
\author{Daniel de Florian}
\affiliation{International Center for Advanced Studies (ICAS), UNSAM, 
Campus Miguelete, 25 de Mayo y Francia (1650) Buenos Aires, Argentina}
\author{Manuel Epele}
\affiliation{ Instituto de F\'{\i}sica La Plata, CONICET - UNLP,
Departamento de F\'{\i}sica,  Facultad de Ciencias Exactas, Universidad de
La Plata, C.C. 69, La Plata, Argentina}
\author{Roger J.\ Hern\'andez-Pinto}
\affiliation{Facultad de Ciencias F\'{\i}sico-Matem\'aticas, 
Universidad Aut\'onoma de Sinaloa, Ciudad Universitaria, CP80000, Culiac\'an, Sinalo, M\'exico}
\author{R. Sassot}
\affiliation{Departamento de F\'{\i}sica and IFIBA,  
Facultad de Ciencias Exactas y Naturales, Universidad de Buenos Aires, Ciudad Universitaria, Pabell\'on\ 1 (1428) Buenos Aires, Argentina}
\author{Marco Stratmann}
\affiliation{Institute for Theoretical Physics, 
University of T\"ubingen, Auf der Morgenstelle 14, 72076 T\"ubingen, Germany}
\begin{abstract}
We revisit the global QCD analysis of parton-to-kaon fragmentation functions 
at next-to-leading order accuracy using the latest experimental information on 
single-inclusive kaon production in electron-positron annihilation, 
lepton-nucleon deep-inelastic scattering, and proton-proton collisions.
An excellent description of all data sets is achieved, and the remaining 
uncertainties in parton-to-kaon fragmentation functions are estimated and discussed
based on the Hessian method. Extensive comparisons to the results from our 
previous global analysis are made.
\end{abstract}
\maketitle

\section{Introduction and Motivation}
%
Parton-to-hadron fragmentation functions (FFs) parametrize
how quarks and gluons that are produced in hard interactions at high energies 
confine themselves into hadrons measured and 
identified in experiment \cite{Feynman:1973xc}. 
This information is beyond the reach of perturbative Quantum Chromodynamics 
(pQCD) and must therefore be inferred from the wealth of data on identified hadron
production under the theoretical assumption that the relevant non-perturbative dynamics 
of FFs factorizes in a universal way from the calculable hard partonic cross sections \cite{ref:fact}
up to small corrections which can be usually neglected.

A precise knowledge on FFs is vital for the quantitative description of 
a wide variety of hard scattering processes designed to probe the spin and flavor structure 
of nucleons and nuclear matter and their interpretation at the most 
elementary and fundamental level. Even though the role of FFs has been highlighted since 
the early days of the parton model \cite{Feynman:1973xc}, only relatively 
recently it has become possible to combine precise enough data from different 
processes with perturbative calculations of matching accuracy to 
determine FFs for identified pions and kaons within meaningful uncertainties
in what is now commonly known as the ``DSS~07 global analysis'' \cite{ref:dss}. 

Precise parton-to-kaon FFs are usually considered as a key ingredient 
to probe the strangeness content of the nucleon and are expected to 
be of crucial importance in further constraining the corresponding momentum distributions
at a future Electron-Ion Collider (EIC) through charged kaon production
in semi-inclusive deep-inelastic scattering (SIDIS) \cite{ref:eic}.
This is especially the case for the helicity-dependent
strangeness parton distributions $\Delta s$ and $\Delta \bar{s}$ \cite{ref:dssv,Aschenauer:2012ve},
largely because of the complete lack of other experimental constraints
from neutrino-induced, electroweak deep-inelastic structure function
measurements \cite{deFlorian:1994wp} that are routinely utilized in all
extractions of unpolarized parton distribution functions (PDFs),
see, e.g., \cite{Ball:2014uwa,Harland-Lang:2014zoa,Dulat:2015mca}.
For the discussions below, it should be kept in mind that
the unpolarized strangeness PDF is also less well constrained 
than the light sea quarks, see, e.g., \cite{Ball:2014uwa}. 

The relatively poor precision achieved for the available 
parton-to-kaon FFs \cite{ref:dss,ref:other-ffs} is an important limiting factor, with relative uncertainties 
roughly one order of magnitude larger than those estimated for the corresponding
parton-to-pion FFs \cite{ref:dss,ref:dss-unc}. This is readily understood
from the fact that pions are much more copiously produced and easier to identify
experimentally than kaons and that their perturbative description is not so much
challenged by potentially large kinematical corrections associated with the hadron's mass 
that is usually neglected in the underlying theoretical framework \cite{Christova:2016hgd}.   
The lack of precision for kaon FFs has, for instance, led to quite some discussions 
\cite{Leader:2014uua} concerning the smallish $\Delta s$ obtained in the analysis of
kaon production in polarized SIDIS \cite{ref:dssv}, a question that can likely only be settled
in the future at an EIC by more precise SIDIS measurements in a broader kinematic range 
along with an improved theoretical analysis based on more reliable kaon FFs.

Since the DSS~07 analysis \cite{ref:dss}, still the only global QCD analysis of kaon FFs available,
strenuous efforts have been made to produce considerably 
more precise data on inclusive hadron production. 
In Ref.~\cite{deFlorian:2014xna}, we performed an update of the DSS~07 results for 
pion FFs (DSS~14), including all the newly available sets of data at that time.
In addition, an iterative Hessian (IH) approach was implemented to assess the uncertainties \cite{ref:ih}
and to provide Hessian uncertainty sets to facility propagating uncertainties related
to FFs to any process of interest.
In what follows, we revisit also our previous global analysis of kaon FFs and perform
similar updates to the latest sets of experimental data and to the way uncertainties are estimated.

More specifically, including single-inclusive electron-positron annihilation (SIA)
data from {\sc BaBar} \cite{ref:babardata} and {\sc Belle} \cite{ref:belledata}
should, in principle, provide a better handle on the gluon-to-kaon FF through QCD 
scaling violations of the SIA structure functions between the scale $Q=M_Z$, 
relevant for the LEP and SLAC experiments included in DSS~07, 
and the scale corresponding to the center-of-mass system (c.m.s.) 
energy of {\sc BaBar} and {\sc Belle}, $Q=\sqrt{S}\simeq 10.5\,\mathrm{GeV}$. 
In addition, since the electroweak couplings of up-type and down-type 
quarks to the $Z$ boson become almost equal at $Q \approx M_Z$, LEP and 
SLAC data are mainly sensitive to the total quark singlet FF for any observed 
hadron $H$. At the lower $\sqrt{S}$ of {\sc BaBar} and {\sc Belle}, 
the quark-antiquark pairs in SIA are produced according to their electrical 
charge, which, in our global fit, should allow for some partial flavor 
separation of kaon FFs.

Another important and new ingredient to the current global analysis 
is the final SIDIS data for proton and deuteron targets 
released by the {\sc Hermes} Collaboration \cite{ref:hermesmult}, 
which supersede the preliminary data 
\cite{ref:hermes-old} utilized (only for proton targets) in the DSS~07 fit. 
This time, we include both the $z-Q^2$ and $z-x$ projections of
the multi-dimensional {\sc Hermes} multiplicities, at variance with 
DSS~07 where only the $z-Q^2$ projections were considered.
Since most of the events [{\cal O}(70\%)] in either of the two projections are 
not shared \cite{Aschenauer:2015rna}, both sets provide highly valuable 
information worth including in the fit.
In addition, since the different bins in $x$ involve 
combinations of the FFs weighted by significantly different PDFs, 
the $z-x$ projections are expected to lead to a much tighter constraint
on the flavor separation than the $z-Q^2$ projections alone, 
where this information is integrated out and potentially diluted.
We also pay special attention to the 
kinematical dependence of the SIDIS cross sections within each 
bin, and integrate these contributions rather than computing the cross 
sections at the mean kinematical values as quoted by the experiment, 
which may lead to significant differences for the estimated multiplicity values 
\cite{Aschenauer:2015rna}.

Another crucial addition to the available suite of data on identified charged
kaons are first multiplicity results in SIDIS from the {\sc Compass} 
experiment at CERN \cite{Adolph:2016bwc}. These data
are very precise despite exhibiting a rather fine binning in the relevant 
kinematic variables. Most importantly, 
{\sc Compass} multiplicities reach much higher values of 
momentum transfer $Q^2 \lesssim 60\,\mathrm{GeV}^2$ than {\sc Hermes} 
$Q^2 \lesssim 30\,\mathrm{GeV}^2$ and, therefore, combining them in
a global fit not only allows us to test and quantify their level of consistency, but
should, in principle, also lead to a considerably better flavor separation of 
the obtained parton-to-kaon FFs.
Not surprisingly, the largest differences with respect to the original DSS~07 
analysis are found mainly at the higher $Q^2$ values not covered by the
{\sc Hermes} data.

Finally, first results on single-inclusive kaon spectra at high transverse 
momenta $p_T$ have become available from the LHC at c.m.s.\ energies of up 
to $2.76\,\mathrm{TeV}$ \cite{ref:alicedata}, which nicely supplement the data 
from BNL-RHIC taken at $\sqrt{S}=200\,\mathrm{GeV}$ that have been already 
used in the original DSS~07 analysis.  Here, we also include new 
results from the {\sc Star} Collaboration for charged kaon production at 
$\sqrt{S}=200\,\mathrm{GeV}$ \cite{ref:starratio11}.

The main goal of our new analysis is to extract an updated, more precise set 
of parton-to-kaon FFs and to determine their uncertainties reliably based 
on the IH method \cite{ref:ih} in light of all the newly available
experimental results in SIA, SIDIS, and $pp$ collisions.
This will allow us to scrutinize the consistency of the information on FFs 
extracted across the different hard scattering processes, i.e., to validate 
the fundamental notion of universality, which is at the heart of any pQCD 
calculation based on the factorization of short- and long-distance physics 
\cite{ref:fact} mentioned at the beginning. 
Since extractions of leading order (LO) FFs have yielded a much less 
satisfactory description of the available pion production data in the 
past \cite{ref:dss}, we only perform our global QCD fit at 
next-to-leading order (NLO) accuracy. 
We note, that first efforts have started recently to perform extractions of
pion FFs from SIA data at next-to-next-to-leading order (NNLO) accuracy \cite{Anderle:2015lqa}
or even by including all-order resummations \cite{Anderle:2016czy}.
Since the relevant cross sections for SIDIS and $pp$ collisions 
are not yet available at NNLO accuracy, a global QCD analysis of FFs can be consistently performed
only at the NLO level for the time being.

The remainder of the paper is organized as follows: in the next section, we briefly 
summarize the main aspects of our updated global analysis, including the choice of
the functional form used to parametrize the FFs at the initial scale $Q_0$ for the QCD 
evolution, the selection of data sets and the cuts imposed on them, and the treatment 
of experimental normalization uncertainties.
The outcome of the new fit is discussed in depth in Sec.~III. 
The obtained parton-to-kaon fragmentation functions and their uncertainties
are shown and compared to the results of our previous global analysis.
Detailed comparisons to the individual data sets are given to demonstrate the 
quality of the fit. Potential open issues and tensions among the different data 
sets will be discussed. We briefly summarize the main results in Sec.~IV.

\section{Technical Framework}
%
Since the main features and technical details of our global QCD 
extractions of FFs for various types of identified hadrons
have already been discussed at length in the literature 
\cite{ref:dss,ref:dss2,ref:eta,ref:dss-unc,deFlorian:2014xna},
we mainly focus in the following on those aspects that differ from the 
original DSS~07 analysis of parton-to-kaon FFs \cite{ref:dss}.
%
\subsection{Functional Form and Fit Parameters \label{sec:funcform}}
%
As in the case of our updated pion FFs in Ref.~\cite{deFlorian:2014xna}, 
the functional form adopted in the original DSS~07 global analysis \cite{ref:dss}  
is still flexible enough to accommodate also the wealth of new experimental 
information included in the present fit. Therefore, we continue to 
parametrize the hadronization of a parton of flavor $i$ into a positively 
charged kaon $K^+$ at an initial scale of $Q_0=1\,\mathrm{GeV}$ as
\begin{equation}
\label{eq:ff-input}
D_i^{K^+}\!(z,Q_0) =
\frac{N_i\, z^{\alpha_i}(1-z)^{\beta_i} [1+\gamma_i (1-z)^{\delta_i}] }
{B[2+\alpha_i,\beta_i+1]+\gamma_i B[2+\alpha_i,\beta_i+\delta_i+1]}\;.
\end{equation}
Here, $B[a,b]$ denotes the Euler Beta-function, and the $N_i$ in 
(\ref{eq:ff-input}) are chosen in such a way that they represent the 
contribution of $z D_i^{K^+}$ to the momentum sum rule.
$z$ is the fraction of momentum of the parton $i$ taken by the kaon.
As in our previous analysis \cite{ref:dss}, we fit $D_{u+\bar{u}}^{K^+}$ 
and $D_{s+\bar{s}}^{K^+}$, containing the ``valence'' quarks in a $K^+$ meson,
independently, but use a single parameterization 
for all the unfavored quark flavors since the data are still unable to 
discriminate between them. Different flavor-breaking scenarios have been explored, 
but they do not change the quality of the fit or even lead to a poor convergence
of the fit due to the extra parameters that need to be introduced.
However, the improved experimental information available in the present fit
now allows us to impose less constraints on the parameter space spanned
by the input function in Eq.~(\ref{eq:ff-input}). Specifically, in
Ref.~\cite{ref:dss} some parameters had to be set to fixed values from the start, 
whereas now all the parameters can exploit a greater degree of flexibility in the fit
and, in principle, can be determined by data. 
Of course, one always has to ensure proper convergence of the fit and, since we
are interested in Hessian uncertainty sets, avoid any parameters that
are only very weakly constrained. Specifically, it turns out that 
$\beta_g\simeq \beta_{\bar{u}}$, $\gamma_{s+\bar{s}}\simeq \gamma_{\bar{u}}$,
and $\delta_{s+\bar{s}}\simeq \delta_{\bar{u}}$, such that we decided to
identify these parameters with each other without any change in the 
total $\chi^2$ of the fit.

No new charm or bottom-tagged data in SIA have become available since the DSS~07 analysis
but the new, very precise results from {\sc BaBar} \cite{ref:babardata} and 
{\sc Belle} \cite{ref:belledata} in SIA and from {\sc Compass} \cite{Adolph:2016bwc} 
and {\sc Hermes} \cite{ref:hermesmult} in SIDIS now constrain both the 
total quark singlet fragmentation function, i.e., summed over all flavors, and the individual, 
flavor-separated light quark FFs much better than before. 
Nevertheless, it turns out that the charm- and bottom-to-kaon FFs can 
still accommodate these changes with $\gamma_{c+\bar{c}}=\gamma_{b+\bar{b}}\simeq 0$
such that we can identify these parameters with zero. In addition, one can
set $\alpha_{c+\bar{c}}=\alpha_{b+\bar{b}}$ without any change to the fit.
As in the DSS~07 and other analyses of FFs, we include heavy flavor FFs 
discontinuously as massless partons in the QCD scale evolution above 
their $\overline{\text{MS}}$-scheme ``thresholds'', $Q=m_{c,b}$,
with $m_c$ and $m_b$ denoting the mass of the charm and bottom quark,
respectively. We note, that the effects of accounting for heavy quark masses 
in extracting light hadron FFs have been explored recently in the case of pion FFs 
with interesting results \cite{Epele:2016gup}. For the time being, and to
allow for a comparison to our previous results from the DSS~07 fit,
we restrict ourselves in the current analysis to the usually adopted 
``zero-mass variable flavor number approximation''.

In total we now have 20 free fit parameters describing our updated FFs for 
quarks, antiquarks, and gluons into a positively charged kaon. They are 
determined from data by a standard $\chi^2$-minimization procedure 
that includes a $\chi^2$-penalty from computing the optimum relative normalization 
of each experimental set of data analytically as was outlined in the DSS~14 analysis 
of pion FFs \cite{deFlorian:2014xna}. 
The latter treatment is at variance with the DSS~07 analysis, where the data sets 
were allowed to float without any $\chi^2$-penalty within the quoted normalization uncertainties. 
The corresponding FFs for negatively charged kaons are obtained, as usual, by charge conjugation
symmetry.

\subsection{Data Selection \label{sec:datasets}}
%
In addition to the data sets already used in the DSS~07 global analysis \cite{ref:dss}, we 
now utilize the new results from {\sc BaBar} \cite{ref:babardata} and {\sc Belle} 
\cite{ref:belledata} in SIA at a c.m.s.\ energy of $\sqrt{S}\simeq 10.5\,\mathrm{GeV}$.
Both sets are very precise and reach all the way up to kaon momentum fractions $z$ close 
to one, well beyond of what has been covered so far by SIA data.
We analyze both sets with $n_f=4$ active, massless flavors using the standard expression 
for the SIA cross section at NLO accuracy. As is customary, we limit ourselves to data with $z\ge 0.1$ 
to avoid any potential impact from kinematical regions where finite, but neglected, hadron 
mass corrections, proportional to $M_{K}/(S z^2)$, might become of any importance. 
Since mass effects grow considerably at lower $\sqrt{S}$ we only use data from {\sc BaBar}
with $z\ge 0.2$; there are no {\sc Belle} data below that cut.
For all previous SIA data, taken at higher $\sqrt{S}$, we use $n_f=5$ and also $z\ge 0.1$, 
following the original DSS~07 analysis. 
Any incompatibility of the two new, precise sets of data at $\sqrt{S}\simeq 10.5\,\mathrm{GeV}$ 
with each other or with the old LEP and SLAC data at $\sqrt{S}\simeq 91.2\,\mathrm{GeV}$ 
\cite{ref:alephdata,ref:delphidata,ref:opaldata,ref:slddata} has the potential to 
seriously spoil the quality of the global fit. 

In case of SIDIS, we replace the preliminary multiplicity data from 
{\sc Hermes} \cite{ref:hermes-old} by their final 
results \cite{ref:hermesmult}. More specifically, we use the data for 
charged kaon multiplicities in four bins of $z$
as a function of both momentum transfer $Q^2$ and the target nucleon's 
(proton or deuteron) momentum fraction $x$.
The kinematical ranges of average values of $Q^2$ and $x$ covered by these data 
are from about $1.1\,\mathrm{GeV}^2$ to $7.4\,\mathrm{GeV}^2$ and $0.064$ to 
$0.277$, respectively, for the $z-Q^2$ projections 
and from about $1.19\,\mathrm{GeV}^2$ to $10.24\,\mathrm{GeV}^2$ and 
$0.034$ to $0.45$, respectively, for the $z-x$ projections,
with $0.2\le z \le 0.8$. Most of the events, {\cal O}(70\%), in either of the two projections 
are not shared \cite{Aschenauer:2015rna}.

In addition, we include for the first time multiplicity data for $K^{\pm}$
production from the {\sc Compass} Collaboration \cite{Adolph:2016bwc}, which are 
given as a function of $z$ in bins of inelasticity $y$ (i.e.\ $Q^2$)
and the initial-state momentum fraction $x$. 
The coverage in $z$ is the same as for the {\sc Hermes} data, but due to the higher 
$\sqrt{S}$ of the {\sc Compass} experiment, the reach in $x$ and $Q^2$ is significantly
broader. Experimental information is available for $0.004\le x \le 0.7$ and 
$1.2 \le Q^2 \le 60\,\mathrm{GeV}^2$.
It turns out, that we do not have to impose any cuts on both data sets to 
accommodate them in the global analysis. As for the SIA data, having now available 
two precise sets of multiplicity data in SIDIS, covering somewhat different but partially 
overlapping kinematics, makes it very important to validate their 
consistency in a global fit.

Finally, we update and add new sets of data for inclusive high-$p_T$ 
kaon production in $pp$ collisions with respect to those included in the 
DSS~07 analysis. 
Most noteworthy are the first results for the kaon-to-pion ratio from the 
{\sc Alice} Collaboration at CERN-LHC \cite{ref:alicedata}, covering 
unprecedented c.m.s.\ energies of up to $2.76\,\mathrm{TeV}$. In addition, we
include {\sc Star} data taken at $\sqrt{S}=200\,\mathrm{GeV}$ for charged kaon 
production and for the $K^-/K^+$ ratio \cite{ref:starratio11}. 
As was discussed in detail in Ref.~\cite{deFlorian:2014xna} in the context of pion FFs, 
it turns out that a good global fit of RHIC and LHC $pp$ data, 
along with all the other world data, can only be achieved if one imposes 
a cut on the minimum $p_T$ of the produced hadron of about $5\,\mathrm{GeV}$. 
We maintain this cut also for the present global analysis, but we will illustrate
how the obtained fit extrapolates to data at lower values of $p_T$. 
Such a $p_T$-cut eliminates all the old $pp$ data sets included in the previous DSS~07 
analysis from the fit, specifically, the {\sc Brahms} \cite{ref:brahmsdata} and the
{\sc Star} \cite{Abelev:2006cs} data.

In Tab.~\ref{tab:exppiontab}, we list all data sets included in our global 
analysis along with the individual $\chi^2$ values obtained in the fit, to which
we now turn.
\begin{table}[thb!]
\caption{\label{tab:exppiontab}Data sets used in our NLO global analysis, 
their optimum normalization shifts $N_i$,
the individual $\chi^2$ values 
(including the $\chi^2$ penalty from the obtained $N_i$), 
and the total $\chi^2$ of the fit. In case of SIDIS, we 
denote the charge $K^{\pm}$, the target hadron (p) or (d), and, for {\sc Hermes},
also the data projection $z-Q^2$ and $z-x$ as $Q^2$ and $x$, respectively. }
\begin{ruledtabular}
\begin{tabular}{lcccc}
experiment& data & norm.  & \# data & $\chi^2$ \\
          & type & $N_i$  & in fit  &          \\ \hline
{\sc Tpc} \cite{ref:tpcdata}  & incl.\ &  1.003 & 12 & 13.4 \\   
{\sc Sld} \cite{ref:slddata}  & incl.\  &  1.014 & 18 & 17.2 \\  
          & $uds$ tag         &  1.014 & 10 & 31.5  \\    
          & $c$ tag           &  1.014 & 10 & 21.3  \\    
          & $b$ tag           &  1.014 & 10 & 11.9  \\    
{\sc Aleph} \cite{ref:alephdata}    & incl.\  & 1.026 & 13 &  29.7 \\ 
{\sc Delphi} \cite{ref:delphidata}  & incl.\  & 1.000  & 12 & 6.9 \\  
                     & $uds$ tag   &  1.000  & 12 & 13.1 \\  
                     & $b$ tag     &  1.000  & 12 & 11.0 \\   
{\sc Opal} \cite{ref:opaleta}  
          & $u$ tag &  0.778  & 5 & 9.6 \\   
          & $d$ tag &  0.778  & 5 & 7.7  \\
          & $s$ tag &  0.778  & 5 & 23.4 \\
          & $c$ tag &  0.778  & 5 & 42.5 \\
          & $b$ tag &  0.778  & 5 & 16.9 \\            
{\sc BaBar} \cite{ref:babardata}     & incl.\ &   1.077 & 45  & 30.6 \\   
{\sc Belle} \cite{ref:belledata}     & incl.\ &   0.996 & 78  & 15.6 \\    \hline  
{\sc Hermes} \cite{ref:hermesmult}  & $K^+$  (p) $Q^2$&  0.843 & 36 & 61.9   \\  
                                    & $K^-$  (p) $Q^2$&  0.843 & 36 & 29.6    \\   
                                    & $K^+$  (p) $x$&  1.135 & 36 &  75.8    \\   
                                    & $K^-$  (p) $x$&  1.135 & 36 & 42.1      \\                                 
                               & $K^+$  (d) $Q^2$&  0.845 & 36 & 44.7         \\  
                               & $K^-$  (d) $Q^2$&  0.845 & 36 & 41.9         \\   
                               & $K^+$ (d)  $x$&  1.095 & 36 &  48.9          \\   
                               & $K^-$ (d) $x$&  1.095 & 36  &  44.4          \\   
{\sc Compass} \cite{Adolph:2016bwc}       & $K^+$ (d)& 0.996 & 309 & 285.8     \\
                                          & $K^-$ (d)& 0.996 & 309 & 265.1     \\  \hline                               

{\sc Star} \cite{ref:starratio11}    & $K^+$,$K^-/K^+$            &  1.088 &  16 & 7.6       \\    
{\sc Alice} \cite{ref:alicedata} \hfill 2.76~TeV& $K/\pi$         & 0.985 &  15 & 21.6        \\ \hline\hline
{\bf TOTAL:} & & & 1194 & 1271.7 \\  
\end{tabular}
\end{ruledtabular}
\end{table}
%

\section{Results}
%
In this section we present and discuss in depth the results of our global
analysis of parton-to-kaon FFs. First, we present the optimum fit parameters,
normalization shifts, and the individual $\chi^2$ values of each data set.
Next, the newly obtained $D_i^{K^{+}}(z,Q^2)$ and their uncertainty estimates
are shown and compared to the results of the previous DSS~07 fit.
The quality of the fit to SIA, SIDIS, and $pp$ data and potential
open issues and tensions among the different sets of data
are illustrated and discussed in Secs.~\ref{sec:sia-data}, \ref{sec:sidis-data},
and \ref{sec:pp-results}, respectively.
%
\subsection{Parton-To-Kaon Fragmentation Functions \label{sec:ff-results}}

In Table~\ref{tab:nlokaonpara} we list the obtained set of parameters in Eq.~(\ref{eq:ff-input})
specifying our updated, optimum parton-to-kaon fragmentation functions at NLO accuracy
at the input scale $Q_0=1\,\text{GeV}$ for the light quark flavors and the gluon,
and for the charm and bottom quarks at their respective mass thresholds $Q_0=m_{c,b}$.
%
%
\begin{table}[th!]
\caption{\label{tab:nlokaonpara}Parameters describing the NLO FFs for positively charged
kaons, $D_i^{K^+}(z,Q_0)$,
in Eq.~(\ref{eq:ff-input}) in the $\overline{\mathrm{MS}}$ scheme at the input scale $Q_0=1\,\mathrm{GeV}$.
Results for the charm and bottom FFs refer to
$Q_0=m_c=1.43\,\mathrm{GeV}$ and
$Q_0=m_b=4.3\,\mathrm{GeV}$, respectively.}
\begin{ruledtabular}
\begin{tabular}{cccccc}
flavor $i$ &$N_i$ & $\alpha_i$ & $\beta_i$ &$\gamma_i$ &$\delta_i$\\
\hline
$u+\overline{u}$                & 0.0663 &-0.486 & 0.098 & 10.85 & 1.826 \\
$s+\overline{s}$                & 0.2319 & 2.745 & 2.867 & 59.07 & 7.421 \\
$\overline{u}=d=\overline{d}=s$ & 0.0059 & 3.657 & 12.62& 59.07 & 7.409 \\
$c+\overline{c}$                &0.1255 & -0.941 & 2.145 & 0.0    & 0.0\\
$b+\overline{b}$                &0.0643 & -0.941 & 5.221 & 0.0    & 0.0\\
$g$                             &0.0283 & 13.60 & 12.62 & 0.0   & 0.0 \\
\end{tabular}
\end{ruledtabular}
\end{table}
%
\begin{figure}[thb!]
\vspace*{-0.4cm}
\begin{center}
\epsfig{figure=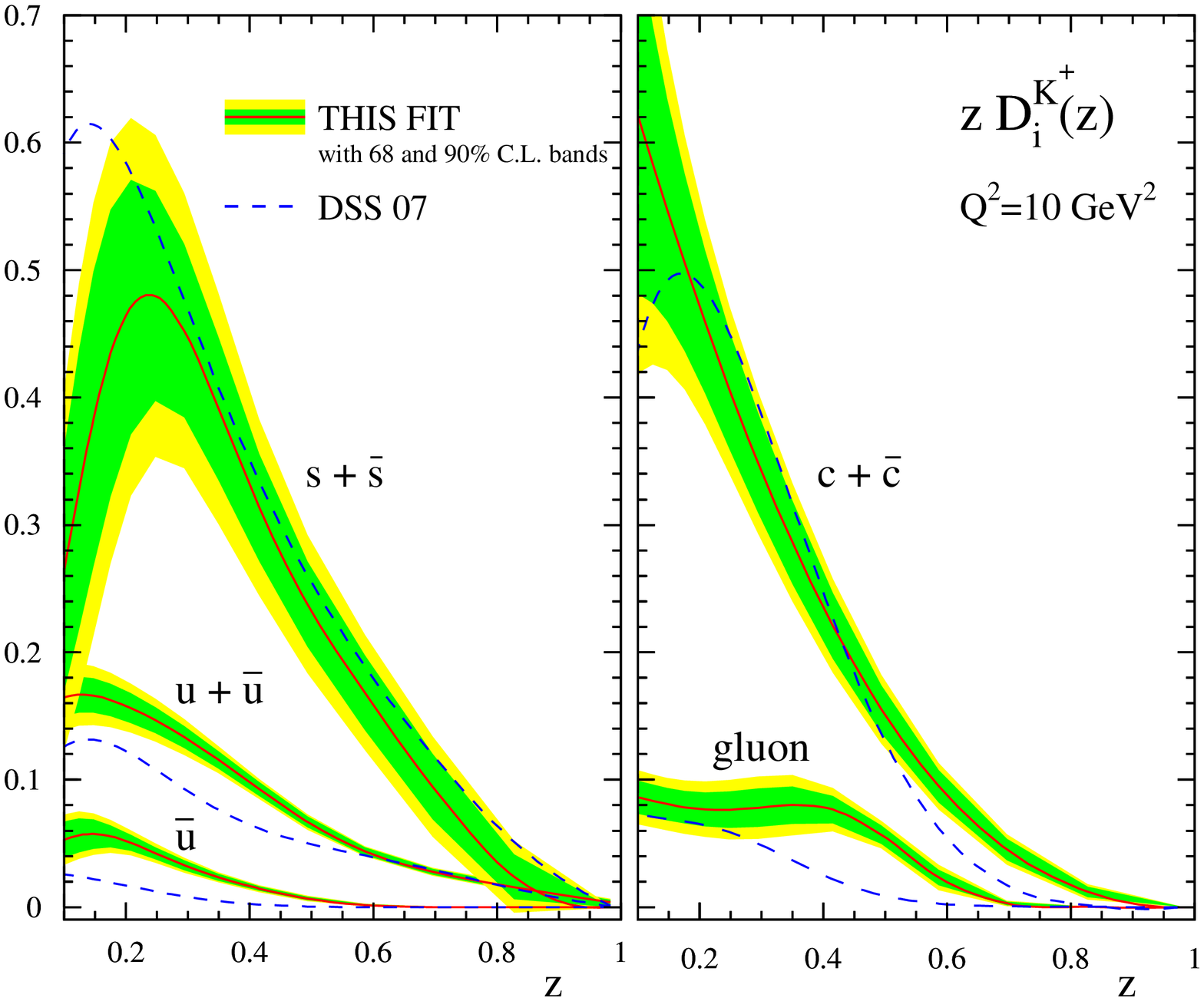,width=0.51\textwidth}
\end{center}
\vspace*{-0.5cm}
\caption{The individual FFs for positively charged kaons $zD_i^{K^{+}}(z,Q^2)$ at
$Q^2=10\,\mathrm{GeV}^2$ (solid lines) along with uncertainty estimates at 
$68\%$ and $90\%$ C.L.\ indicated by the inner and outer shaded bands, respectively.
Also shown is a comparison to our previous DSS~07 global analysis \cite{ref:dss} 
(dashed lines).
\label{fig:ff-at-10}}
%
\vspace*{-0.4cm}
\begin{center}
\epsfig{figure=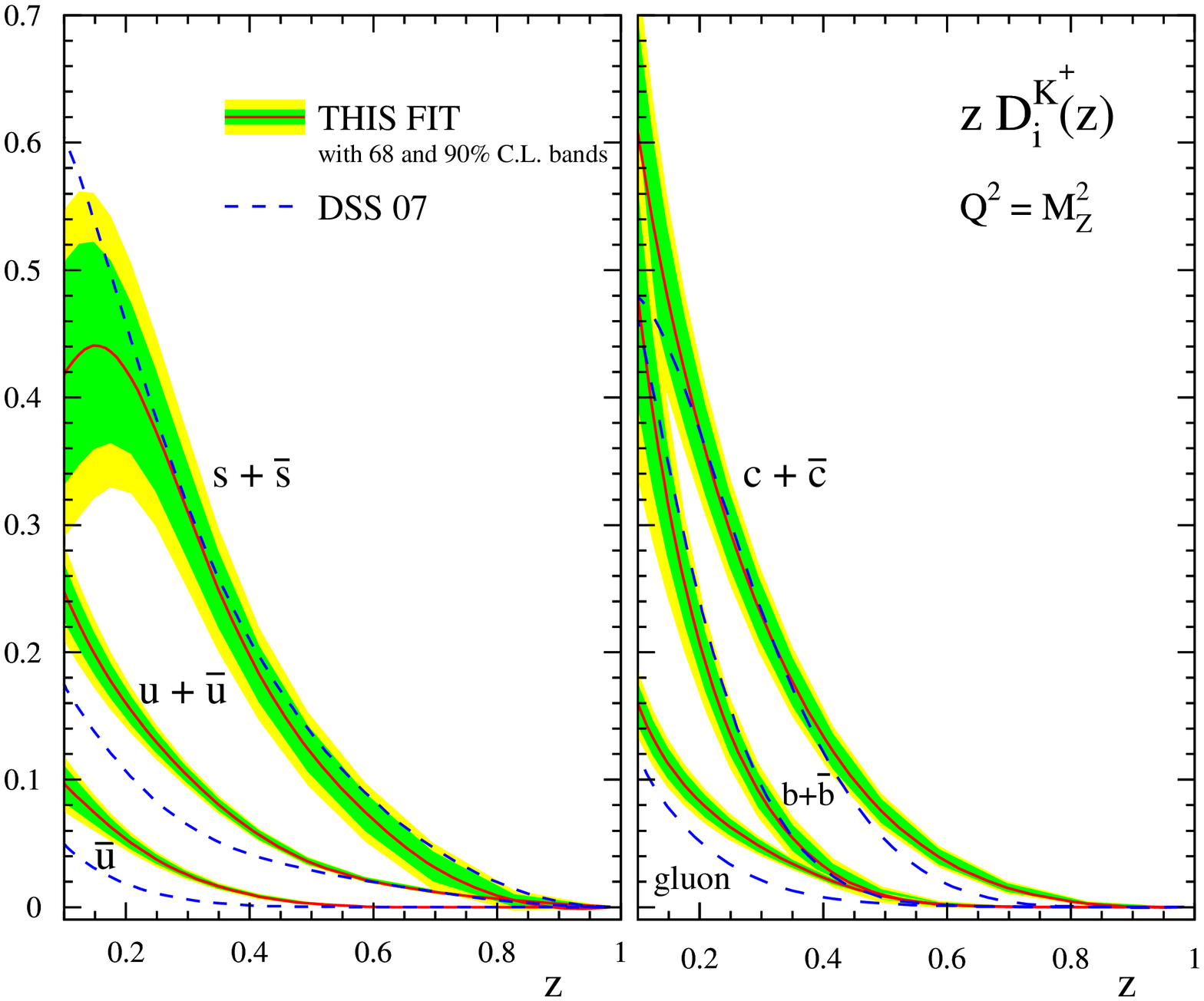,width=0.51\textwidth}
\end{center}
\vspace*{-0.5cm}
\caption{As in Fig.~\ref{fig:ff-at-10} but now at the scale $Q=M_Z$ where also
the bottom-to-$K^+$ fragmentation function is nonzero.
\label{fig:ff-at-mz}}
\vspace*{-0.2cm}
\end{figure}
The new NLO FFs $D_i^{K^+}(z,Q^2)$, evolved to two different values of $Q^2$,
are shown as a function of $z$ in Figs.~\ref{fig:ff-at-10} and \ref{fig:ff-at-mz}
along with our estimates of uncertainties at both $68\%$ and $90\%$ confidence level (C.L.)
and the results from our previous DSS~07 fit \cite{ref:dss}.
As can be inferred from the figures, the FFs for most flavors are either close
to the updated fit or within its $90\%$ C.L.\ uncertainty band; one should recall, that
only data with $z\ge 0.1$ are included in our analysis [$z\ge 0.2$ for {\sc{BaBar}}].
For some flavors $i$ and regions of $z$ there are, however, sizable differences.
They are most noticeable for $D_{u+\bar{u}}^{K^{+}}$ and the unfavored
FF $D_{\bar{u}}^{K^{+}}$ below $z\simeq 0.5$, for $D_{c+\bar{c}}^{K^{+}}$
at large $z$, and for the gluon-to-kaon FF around $z\simeq 0.4$. 

The differences with respect to the DSS~07 results are mainly driven by the newly added 
{\sc Belle} and {\sc Babar} data at high $z$, by the $z-x$ projections of the multiplicities 
both from {\sc Hermes} \cite{ref:hermesmult} and {\sc Compass} \cite{Adolph:2016bwc}, 
and by the $K^-/K^+$ ratios measured in $pp$ collisions by {\sc Star} \cite{ref:starratio11}.
All these sets provide sensitivity to the flavor separation of the parton-to-kaon FFs
that was not available in the DSS~07 analysis, and in the global fit 
all FFs have to adjust accordingly.
It is worth noticing that the total strange quark FF
$D_{s+\bar{s}}^{K^{+}}$, which plays an important role in determinations of
the strangeness helicity distribution \cite{ref:dssv}, is always somewhat 
smaller than the corresponding DSS~07 result, but the differences 
are within the $90\%$ C.L.\ uncertainty band for $z\gtrsim 0.1$. 
In spite of the much improved experimental information, no evidence of a 
flavor symmetry breaking between the unfavored FFs is found. A single parameterization for 
$D^{K^+}_{\overline u}=D^{K^+}_d=D^{K^+}_{\overline d}=D^{K^+}_s$ is still the 
most economical choice to reproduce the data, as was the case in the original DSS~07 analysis.  

In terms of uncertainties, the strange quark FF is less well constrained than other FFs
despite being a ``favored'' FF. Light quark FFs have the advantage that 
$u$ and $d$ quarks are much more abundant than $s$ quarks in SIDIS due to the corresponding
$u$ and $d$ valence quark PDFs. In addition, scattering off a $u$-quark is more likely due to its
larger electrical charge. The heavy quark FFs are rather tightly constrained by
flavor-tagged SIA data and, thanks to the new {\sc Belle} and {\sc Babar} data,
to some extent also from their interplay with LEP and SLAC data at higher c.m.s.\ energies;
for instance, for {\sc Belle} and {\sc Babar} the bottom FFs does not play a role.

The overall quality of the fit is summarized in Tab.~\ref{tab:exppiontab}, 
where we list all data sets included in our global analysis, as discussed in 
Sec.~\ref{sec:datasets}, along with their individual $\chi^2$ values and the 
analytically determined normalization shifts for each set.
We note that the quoted $\chi^2$ values are based only on fitted data points,
i.e., after applying the cuts mentioned in Sec.~\ref{sec:datasets}, 
and include the $\chi^2$ penalty from the normalization shifts; see
Ref.~\cite{deFlorian:2014xna} for more details on the method.

It is also worth mentioning that there is a more than five-fold increase in the
number of available data points as compared to the original DSS~07 analysis \cite{ref:dss}. 
Secondly, the overall quality of the global fit has improved dramatically 
from $\chi^2/{\mathrm{d.o.f.}}\simeq 1.83$ for DSS~07, see Tab.~V in Ref.~\cite{ref:dss}, 
to $\chi^2/{\mathrm{d.o.f.}}\simeq 1.08$ for the current fit. 
A more detailed inspection reveals that the individual $\chi^2$ values for
the SIA data \cite{ref:alephdata,ref:delphidata,ref:opaldata,ref:slddata,ref:tpcdata},
which were already included in the DSS~07 fit, have, by and large, not changed significantly. 
The biggest improvement concerns the SIDIS multiplicities from {\sc Hermes} 
which, in their published version \cite{ref:hermesmult}, are described rather well by the 
updated fit, with only a few exceptions; see below.
Also, the charged kaon multiplicities from {\sc Compass} \cite{Adolph:2016bwc}
and the new SIA data from {\sc BaBar} \cite{ref:babardata} and {\sc Belle} \cite{ref:belledata}
integrate very nicely into the global QCD analysis of parton-to-kaon FFs at NLO accuracy. 
We recall that the original DSS~07 fit was based on the 2003 NLO (2002 LO) 
PDF set \cite{ref:mrst} (\cite{ref:mrstlo}) from the MRST group. 
In the present fit, the underlying set of PDFs has have been 
upgraded to the recent MMHT~2014 analysis \cite{Harland-Lang:2014zoa}, that gives 
a much more accurate description of sea-quark parton densities on which the analysis of SIDIS 
multiplicities depends rather strongly. We have checked that very similar results for kaon FFs
are obtained with other up-to-date sets of PDFs such as \cite{Ball:2014uwa,Dulat:2015mca}. 
Nevertheless, the corresponding PDF uncertainty is included in the $\chi^2$-minimization procedure
and, hence, the quoted $\chi^2$ values for SIDIS multiplicities.

\subsection{Electron-Position Annihilation Data\label{sec:sia-data}}
\begin{figure}[bht!]
\begin{center}
\epsfig{figure=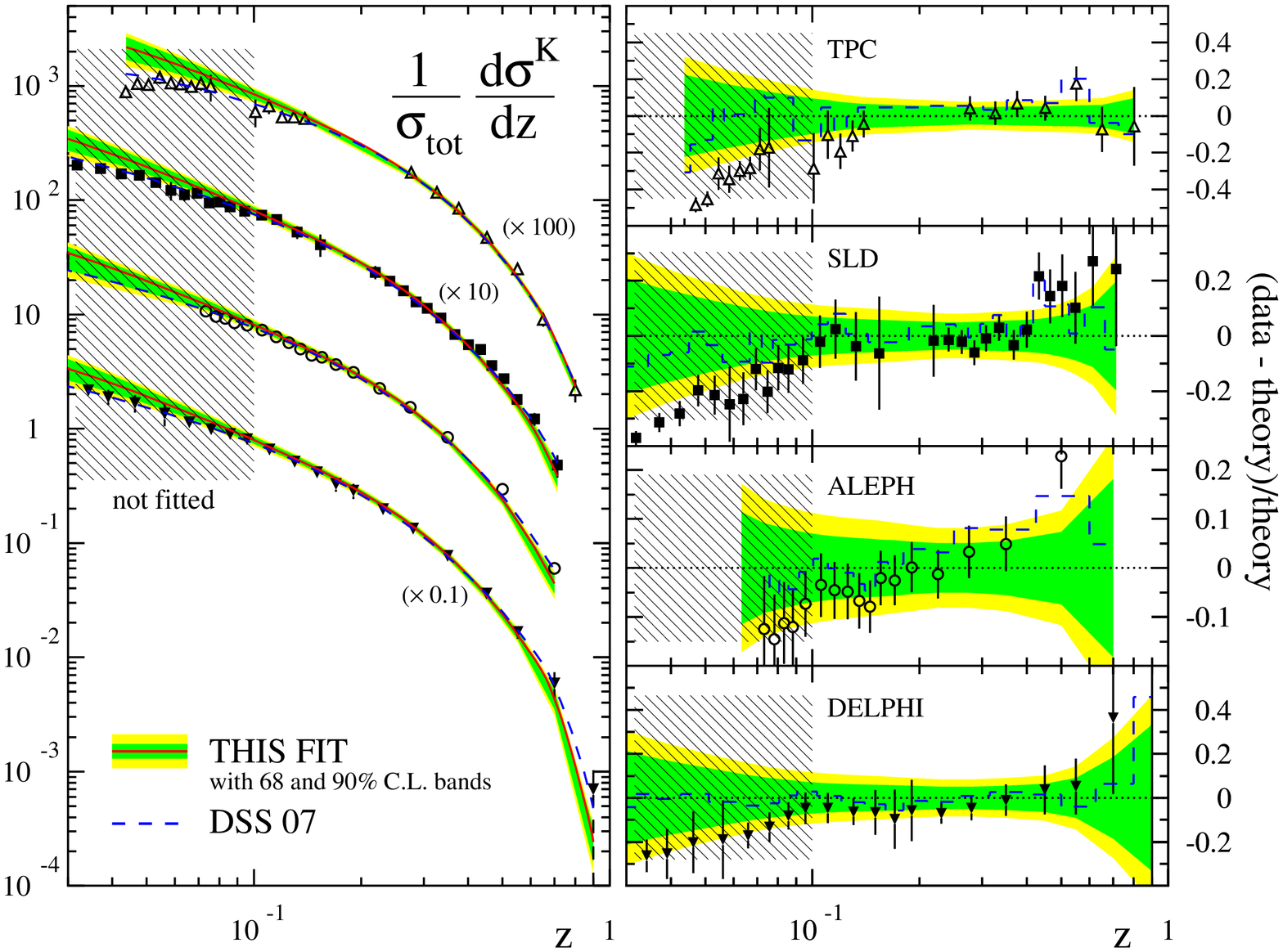,width=0.5\textwidth}
\end{center}
\vspace*{-0.5cm}
\caption{Left-hand side: comparison of our new NLO results (solid lines) 
and the previous DSS~07 fit \cite{ref:dss} (dashed lines) with data sets for 
inclusive kaon production in SIA used in both fits, see Tab.~\ref{tab:exppiontab}.
The inner and outer shaded bands correspond to new uncertainty estimates 
at $68\%$ and $90\%$ C.L., respectively.
Right-hand side: ``(data-theory)/theory'' for each of the data sets w.r.t.\ 
our new fit (symbols) and the DSS~07 analysis (dashed lines).
\label{fig:ee-untagged}}
%
\vspace*{-0.5cm}
\begin{center}
\epsfig{figure=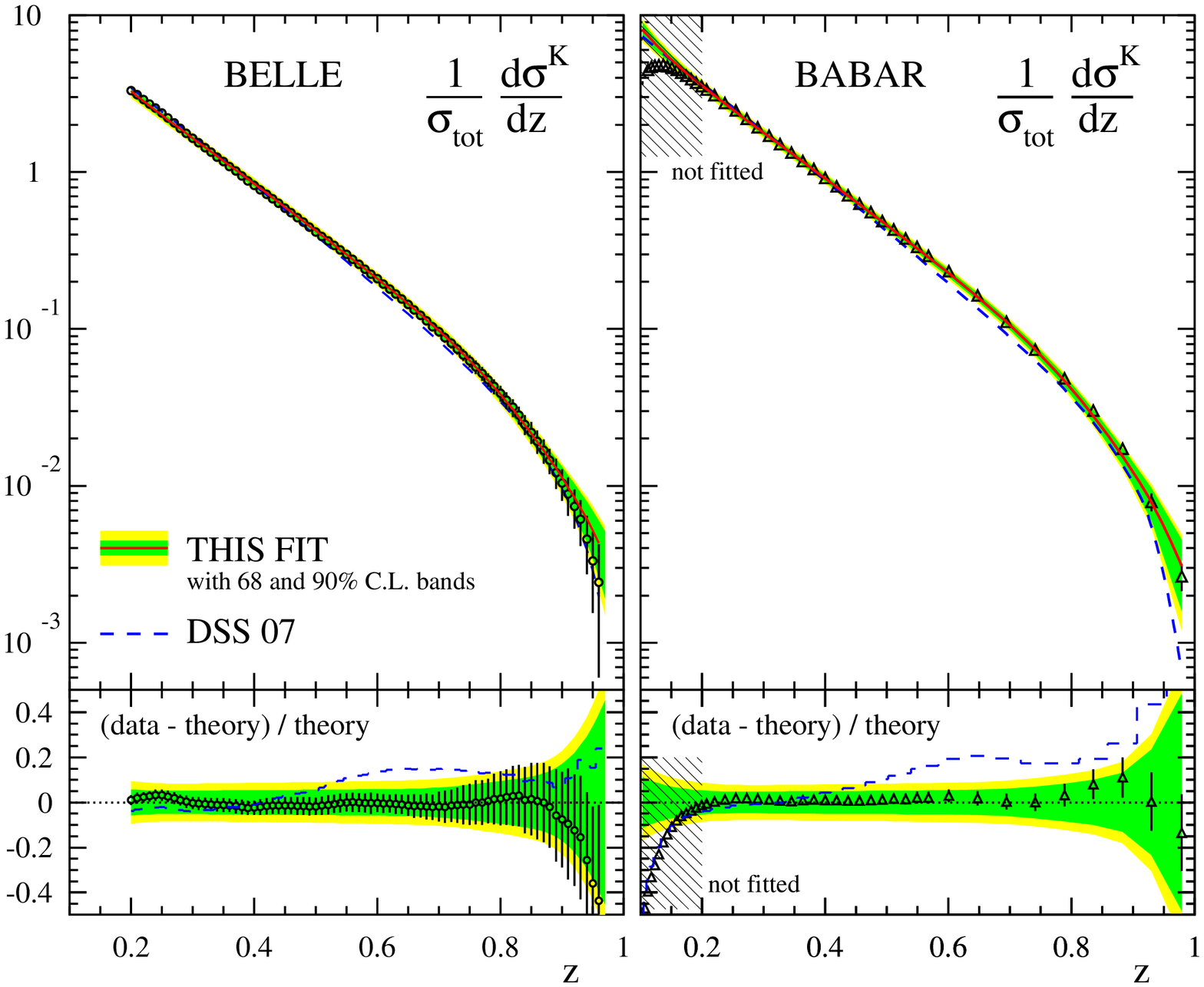,width=0.5\textwidth}
\end{center}
\vspace*{-0.7cm}
\caption{Left-hand side: comparison of our new NLO results (solid line) 
with the new {\sc Belle} data \cite{ref:belledata}; also
shown is the result obtained with the DSS~07 fit \cite{ref:dss} (dashed line).
Right-hand side: same, but now for the {\sc BaBar} data \cite{ref:babardata}.
Lower panels: ``(data-theory)/theory'' for each of the data sets w.r.t.\ our 
new fit (symbols) and the DSS~07 analysis (dashed lines).
The inner and outer shaded bands correspond to the new uncertainty estimates at 
$68\%$ and $90\%$ C.L., respectively. Data below $z=0.2$ are not included in the fit.
\label{fig:ee-belle-babar}}
\end{figure}

%
In Figs.~\ref{fig:ee-untagged} and \ref{fig:ee-belle-babar} we present
a detailed comparison of the results of our fit and its uncertainties at both $68\%$ 
and $90\%$ C.L.\ with the SIA data already included and newly added to the original 
DSS~07 analysis \cite{ref:dss}, respectively. In general, the agreement of the fit with 
SIA data is excellent in the entire energy and $z$-range covered by the experiments.
The new fit reproduces SLAC and LEP data at $Q=M_Z$ as well or even slightly 
better than the old DSS~07 result for $z\ge 0.1$, and improves very significantly the description 
of the newly added {\sc Belle} and {\sc BaBar} data as can be best seen from
the ``(data-theory)/theory'' panels in Fig.~\ref{fig:ee-belle-babar}; recall that 
only data with $z\ge 0.2$ are included in the fit for {\sc BaBar} due the lower $\sqrt{S}$.
This is mainly achieved by changing the singlet flavor combinations 
rather significantly at large $z\sim 0.5 - 0.8$ at the lower $Q$ 
relevant for {\sc Belle} and {\sc BaBar}.
For SIA data at $z$-values lower than those included in the $\chi^2$ minimization, 
the old DSS~07 fit gives, however, a better description when extrapolated, presumably because
the fit has to accommodate much less data.
%

The {\sc Belle} data \cite{ref:belledata}, shown in Fig.~\ref{fig:ee-belle-babar},
provide not only the finest binning in $z$ but also reach the
highest $z$-values measured so far. Above $z\gtrsim 0.8$ one observes an increasing
trend for the new fit to overshoot the data, still within the estimated and growing 
theoretical (and experimental) uncertainties though. In this kinematic regime 
one expects large logarithmic corrections, which appear in each order of perturbation theory,
to become more and more relevant. 
It is known how to resum such terms to all orders in the strong coupling \cite{ref:resum}, 
and it might be worthwhile to explore their relevance 
and whether they could further improve the agreement with data
in a future dedicated analysis in detail. Resummations also provide 
a window to non-perturbative contributions to the perturbative series so far little 
explored. The $z$-binning of {\sc BaBar} data \cite{ref:babardata} is more sparse towards 
large $z$, and a similar trend as for the {\sc Belle} data is not visible here.
%
\begin{figure*}[th!]
\vspace*{-0.4cm}
\begin{center}
\epsfig{figure=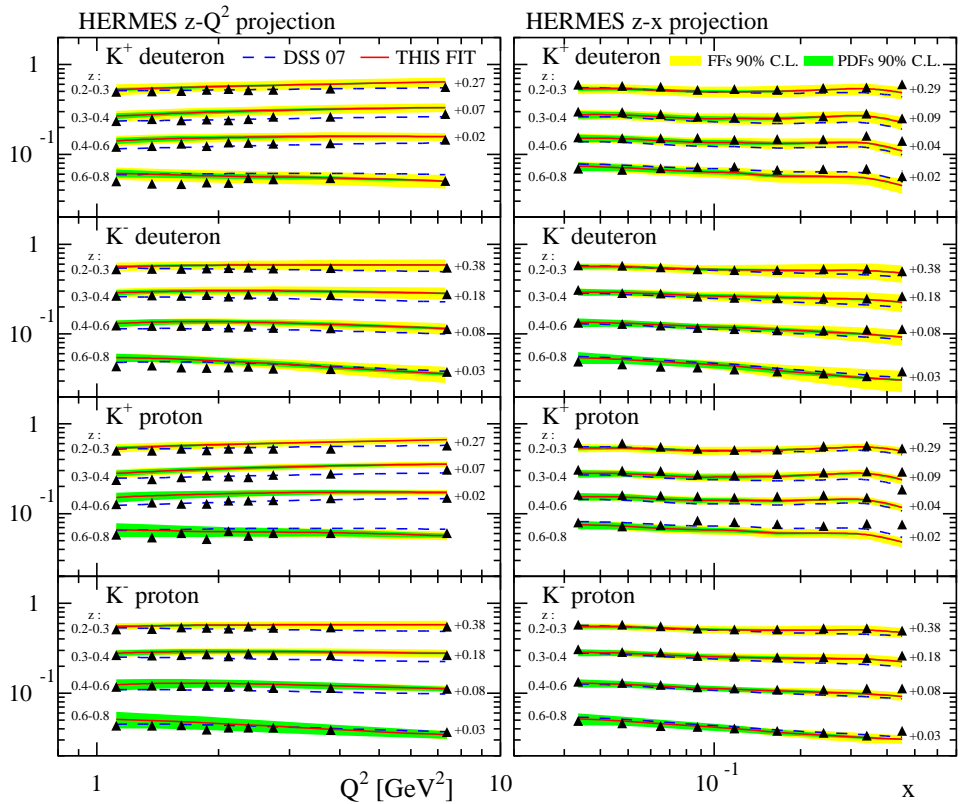,width=0.8\textwidth}
\end{center}
\vspace*{-0.9cm}
\caption{Comparison of our NLO results (solid lines) for
charged kaon multiplicities in SIDIS in four different bins of $z$ 
with data from the {\sc{Hermes}} Collaboration \cite{ref:hermesmult} for both
proton and deuteron targets.
The $z-Q^2$ and $z-x$ projection are shown on the left-hand- and right-hand-side,
respectively. The light shaded (yellow) bands correspond to uncertainty estimates at $90\%$ C.L., 
the dark shaded (green) bands illustrate the PDF uncertainty also at $90\%$ C.L.  
The results obtained with the DSS~07 set of FFs (dashed lines) are included as well. 
As indicated in the plot, different constant factors are added 
to the multiplicities to better distinguish the results for 
different ranges of $z$ in each panel. \label{fig:sidis-hermes}}
\end{figure*} 

Our estimated uncertainty bands at $68\%$ and $90\%$ C.L.\ are also
shown in Figs.~\ref{fig:ee-untagged} and \ref{fig:ee-belle-babar}.
They reflect the accuracy and kinematical coverage of the fitted data sets
and, hence, increase towards both small and large values of $z$, very
similar to the pattern observed for the individual FFs $D_i^{K^{+}}$ 
in Figs.~\ref{fig:ff-at-10} and \ref{fig:ff-at-mz}.
One should keep in mind, however, that the obtained bands are constrained by the fit 
to the {\em global} set of SIA, SIDIS, and $pp$ data and do not necessarily 
have to follow the accuracy of each {\em individual} set of data.

Turning again to the extrapolation of the new fit to data below the cut of $z=0.1$
($z=0.2$ for {\sc BaBar}),
one can infer from Figs.~\ref{fig:ee-untagged} and \ref{fig:ee-belle-babar}
that the NLO theory estimates continue to rise while the data start to drop 
towards lower values of $z$ for all sets.
Such an effect is not unexpected and signifies the onset of 
neglected hadron mass effects in the theoretical framework. 
In fact, this is, in general, the very reason for all fits of FFs to SIA data to choose some lower
cut on $z$. Due to the higher mass of the kaons, we decided to 
impose also a slightly higher cut of $z\ge 0.1$ than what was used 
for pions in the latest DSS~14 analysis \cite{deFlorian:2014xna}.
Another issue with the small-$z$ region has to do with strongly enhanced terms in the
perturbative series of both the time-like splitting functions and the SIA coefficient
functions that become more and more severe in higher orders and would require the use
of all-order resummation techniques \cite{Anderle:2016czy}.
 
\subsection{Semi-Inclusive DIS Multiplicities \label{sec:sidis-data}}
%
The most powerful constraint for flavor-separated FFs comes from charged kaon
multiplicities in SIDIS. Contrary to SIA, which produces $K^+$ and $K^-$ at
equal rates, SIDIS multiplicities are sensitive to the produced hadron's charge,
through the choice of the target hadron and the kinematics, i.e., the
parton momentum fraction $x$ that is probed.
For instance, data taken on a proton target in the valence region (medium-to-large
values of $x$) will contain more $K^+$ than $K^-$ mesons, since $u$-quarks are much 
more abundant in a proton than $\bar{u}$-quarks. 
The access to precisely measured multiplicities for 
positively and negatively charged kaons, produced alternatively off proton 
or deuteron targets, and within different regions of proton momentum fractions $x$
allows for studying the flavor dependence of parton-to-kaon FFs at an
unprecedented level of accuracy. 
%
\begin{figure*}[ht!]
\vspace*{-0.5cm}
\begin{center}
\epsfig{figure=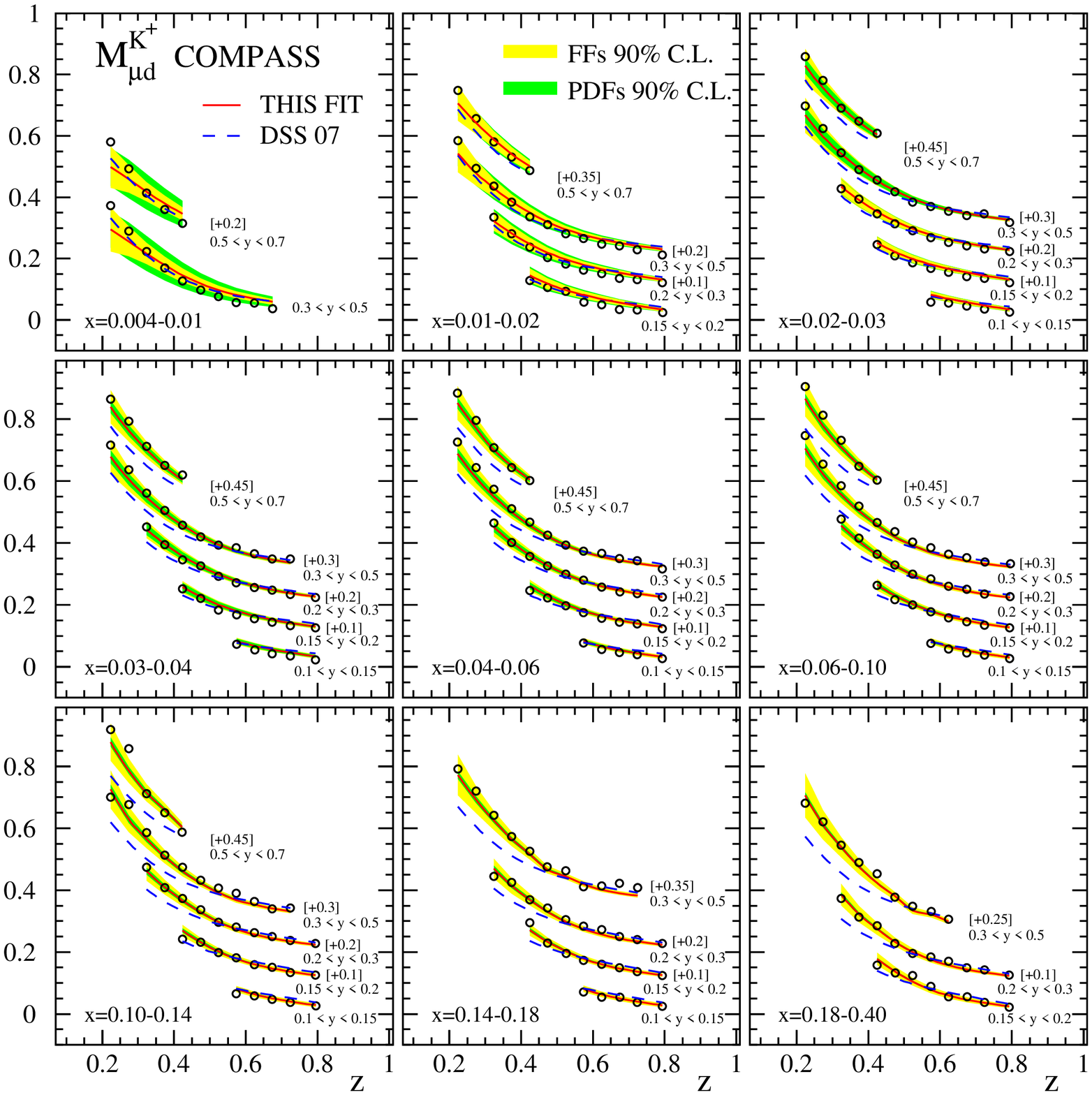,width=0.8\textwidth}
\end{center}
\vspace*{-0.75cm}
\caption{Comparison of our NLO results for $K^+$ multiplicities in SIDIS (solid lines) with data
from the {\sc{Compass}} experiment \cite{Adolph:2016bwc} taken on a deuteron target
for various bins in $x$ and $y$. As in Fig.~\ref{fig:sidis-hermes},
the light (yellow) and dark (green) shaded bands correspond to FF and PDF 
uncertainty estimates at $90\%$ C.L., respectively.
Also shown are the results obtained with our previous DSS~07 set of FFs (dashed lines).
As indicated in the plot, different constant factors are added 
to the multiplicities $M_{\mu d}^{K^{+}}$ to more easily distinguish the results for 
different values of $y$ (i.e.\ $Q^2$) in the same bin of $x$.
\label{fig:sidis-compass-piplus}}
\end{figure*} 

Compared with the DSS~07 analysis, where we only had some preliminary set of
kaon multiplicities on a proton target from the {\sc Hermes} Collaboration 
at our disposal \cite{ref:hermes-old}, we can now use their published, 
final set of data for both proton and deuteron targets \cite{ref:hermesmult}. 
In Fig.~\ref{fig:sidis-hermes} we illustrate the quality of the new fit with respect to
the {\sc Hermes} data. Shown are the charged kaon multiplicities $M_{e,p(d)}^{K^{\pm}}$, 
which are defined as the ratio of the inclusive kaon yield and the total DIS cross 
section in the same $x$ and $Q^2$ bin in lepton-proton ($lp$) or lepton-deuteron 
($ld$) scattering:
\begin{equation}
\label{eq:mult}
M_{l,p(d)}^{K^{\pm}} \equiv \frac{ d\sigma^{K^{\pm}}_{l,p(d)}/dx\,dQ^2\,dz}
{d\sigma_{l,p(d)}/dx\,dQ^2}\;.
\end{equation}

In the global fit we consider the two-dimensional projections of the three-dimensional multiplicity data 
\cite{ref:hermesmult} onto the $z-Q^2$ dependence (left-hand-side of Fig.~\ref{fig:sidis-hermes}) and, for the
first time, also the $z-x$ dependence (right-hand-side), 
for four different bins of the kaon's momentum fraction $z$. The interplay of
these data, i.e., the attempt to fit them both simultaneously,
provides a much improved sensitivity to the flavor-structure of the parton-to-kaon FFs.
As was mentioned in the Introduction and noticed in Ref.~\cite{Aschenauer:2015rna},
we find it important to carefully include the full kinematic dependence of the SIDIS cross section 
by integrating within the boundaries of each bin rather than simply adopting the quoted mean values
for $x$, $Q^2$, and $z$.
 
\begin{figure*}[ht]
\vspace*{-0.5cm}
\begin{center}
\epsfig{figure=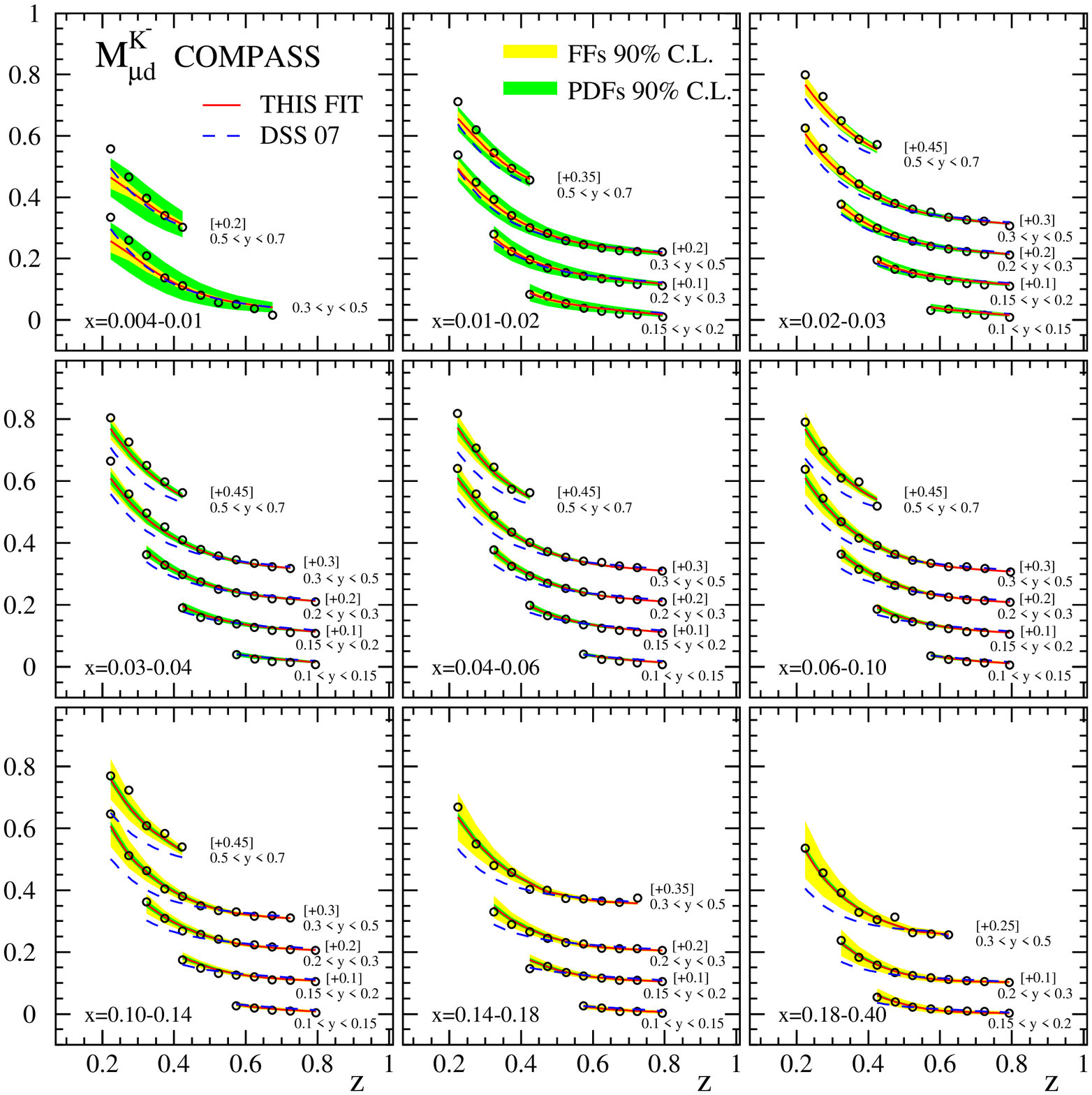,width=0.8\textwidth}
\end{center}
\vspace*{-0.75cm}
\caption{Same as in Fig.~\ref{fig:sidis-compass-piplus} but now for the $K^-$ multiplicities
$M_{\mu d}^{K^{-}}$ .
\label{fig:sidis-compass-piminus}}
\end{figure*} 
The actual extraction of the FFs from SIDIS multiplicities receives a further complication from
the need to choose a certain set of non-perturbative PDFs for the proton (deuteron) target which also
need to be obtained from global QCD analyses to data; 
see, for instance, Refs.~\cite{Ball:2014uwa,Harland-Lang:2014zoa,Dulat:2015mca}.
Much progress has been made in recent years in carefully quantifying and reducing 
the uncertainties of PDFs, thanks to the high demand for precise theoretical estimates
for the CERN-LHC program. Here, and for the analyses of $pp$ data below, 
we adopt the recent set by the MMHT~2014 group \cite{Harland-Lang:2014zoa},
along with their Hessian uncertainty sets to properly propagate PDF uncertainties
to our extraction of FFs, see below. We note that the $\chi^2$ estimates account for the
error inherent to the computation of SIDIS multiplicities arising from varying the input
PDFs within their quoted uncertainties.

Even though the nuclear modification of the PDFs for nucleons
bound in deuteron targets has been acknowledged to be a non-negligible, 
percentish-level effect a long time ago \cite{Epele:1991np},
most quantitative estimates still suggest a flavor-independent, multiplicative factor
that would cancel in the SIDIS multiplicities (except for the small NLO correction
stemming from initial-state gluons), and, therefore, we disregard
this small correction in our analyses. In any case, the fairly sizable 
PDF uncertainties can be viewed to cover for this type of uncertainty as well.

We use the standard Mellin technique \cite{ref:mellin2,ref:dssv} to pre-calculate look-up tables 
for each data point at NLO accuracy to speed up the fitting procedure and to facilitate 
the uncertainty estimates significantly. We recall that at NLO, the relevant hard 
scattering coefficient functions for SIDIS \cite{ref:sia-nlo,ref:sidis-nlo,ref:lambda-nlo} depend 
in a non-trivial way on both $x$ and $z$, such that an often used, naive approximation,
where the $x$ and the $z$ dependence in Eq.~(\ref{eq:mult}) is assumed to completely 
factorize, is inadequate and bound to fail. Even at LO accuracy such an assumption cannot work 
as soon as different quark flavors fragment differently into the observed hadron, which they do 
for charged kaons and all other hadrons.

Before discussing the results, we remark that the use of the {\sc Hermes} multiplicity 
data as a means of providing a reliable flavor and charge separation for kaon FFs 
in the DSS~07 fit was often questioned in the past \cite{ref:elliot} in rather harsh terms; 
likewise, for pion FFs. It has even been suggested that the DSS~07 FFs were 
``inadequate'' since they were based on preliminary data, significantly 
different from the final ones, and, in any case, could not reproduce the multiplicities 
as a function of $x$ \cite{ref:elliot}. 
From Fig.~\ref{fig:sidis-hermes} it is evident that {\em both} {\sc Hermes} projections,
$z-Q^2$ and $z-x$, can be reproduced very well within the estimated uncertainties 
by the present global fit that also includes the {\sc Compass} SIDIS data which cover
a broader range in $Q^2$; see below. Even the old DSS~07 fit (dashed lines), based on a much 
reduced set of data and outdated PDFs, nicely reproduces not only the final $z-Q^2$ dependent
multiplicities by {\sc Hermes} but also the $x-z$ dependent ones, not included 
in the original fit. 

In Fig.~\ref{fig:sidis-hermes} we also show our uncertainty estimates 
at $90\%$ C.L.\ for both the FFs obtained from the fit and 
the MMHT PDFs computed from the corresponding Hessian sets \cite{Harland-Lang:2014zoa}. 
It is worth noticing that depending on the charge of the final-state kaon, 
the type of target, and the kinematics, one or the 
other source of uncertainty prevails, showing what type of data and 
kinematics may help to constraint either FFs or PDFs in the future.  
It also suggests that the averages over charges and/or kinematic bins, 
that is sometimes applied to the multiplicity data, actually dilutes 
their constraining power and may potentiate the propagation of uncertainties.

The newly available data from the {\sc Compass} Collaboration \cite{Adolph:2016bwc}, 
taken at a higher c.m.s.\ energy than the {\sc Hermes} data, 
are a very important ingredient for the present analysis as they shed light on the validity 
of using a standard, leading-twist pQCD framework at NLO accuracy to describe 
multiplicity data for charged kaons at the comparatively low scales $Q^2$ 
reached at {\sc Hermes}.
Achieving a good global fit of data taken at different energies 
and kinematic ranges with a universal set of parton-to-kaon FFs
cannot be taken for granted and provides a non-trivial check for the
consistency of different measurements.

More specifically, in the present fit we include the charged kaon 
results from {\sc Compass} obtained on a deuteron target \cite{Adolph:2016bwc}. 
The data are presented as a function of $z$ in 9 bins of $x$, each subdivided into 
various bins in $y$ that effectively select different $Q^2$-ranges. 
In total 309 data points pass our cuts for both $K^+$ and $K^-$ multiplicities.
The comparison of the {\sc Compass} data to the results of our global
analysis at NLO accuracy 
is presented in Figs.~\ref{fig:sidis-compass-piplus} and \ref{fig:sidis-compass-piminus}.
A very satisfactory agreement is achieved in almost all bins across the entire 
kinematic regime covered by data as can be also inferred from Tab.~\ref{tab:exppiontab};
the obtained $\chi^2/{\mathrm{d.o.f.}}$ for both $K^+$ and $K^-$ multiplicities is about 1.
As in Fig.~\ref{fig:sidis-hermes}, the shaded bands illustrate our uncertainty
estimates at $90\%$ C.L.\ for both the FFs and the PDFs.

First and foremost, these results demonstrate that the low-energy {\sc Hermes} \cite{ref:hermesmult} 
and the new {\sc Compass} \cite{Adolph:2016bwc} charged kaon multiplicity data 
can be described simultaneously and, equally important, without spoiling the agreement 
with SIA results discussed before. 
This is, to a somewhat lesser extent, even the case when one adopts the old DSS~07 set
of kaon FFs. As can be seen from Figs.~\ref{fig:sidis-compass-piplus} and \ref{fig:sidis-compass-piminus}, 
they lead to a fair agreement with the {\sc Compass} data without any re-fitting
except for some of the bins corresponding to the highest $Q^2$ values; 
for the $z-Q^2$ projections of the {\sc Hermes} data, shown in the left panel of
Fig.~\ref{fig:sidis-hermes}, the DSS~07 FFs even lead to a slightly better description of the
data than the new, updated global fit. The bottom line is, that the new {\sc Compass} data 
mainly correct the charge and flavor separation provided by the DSS~07 set of FFs 
at higher $Q^2$ values, an information that was beyond the reach of the {\sc Hermes} data
adopted in the DSS~07 fit. 

We end the discussion of the SIDIS data by noticing that
the $\chi^2/{\mathrm{d.o.f.}}$ values obtained for some of the {\sc Hermes} data, in particular, for
$K^+$ multiplicities on a proton target, are higher than for {\sc Compass}.
While such fluctuations in $\chi^2$ are perfectly normal in a global QCD analysis of
a large amount of data sets, we believe that there is further room for improvement by exploring 
in more detail the interplay of the used set of PDFs with the quality of the fit to SIDIS data. 
None of the available sets of PDFs is constrained by data in most of the kinematic regime
accessible at {\sc Hermes}, mainly because stringent cuts on $Q^2$ are applied in these fits.
Hence, it might be very worthwhile to perform a combined, i.e., simultaneous, 
global analysis of FFs and PDFs in the future, a task which is, however, well beyond the scope
of this paper. A first attempt in this direction will be made in a forthcoming publication \cite{ref:borsa}.

\subsection{RHIC and LHC Data \label{sec:pp-results}}
%
The third and final main ingredient in our global analysis of parton-to-kaon FFs
is the experimental information coming from hadron-hadron collisions,
more specifically, single-inclusive high-$p_T$ kaon production in $pp$ 
collisions at BNL-RHIC and the CERN-LHC.
Compared to the original DSS~07 analysis \cite{ref:dss}, which made use of the 
{\sc Brahms} data for charged kaons and {\sc Star} results for 
$K^0_S$ production at mid rapidities, both sets are limited to
very low values of transverse momentum ($p_T<5\,\mathrm{GeV}$), we can now utilize 
new results from the {\sc Star} Collaboration for charged kaons $K^{\pm}$ 
\cite{ref:starratio11} up to $p_T\simeq 13\,\mathrm{GeV}$ 
as well as first data from the {\sc Alice} Collaboration at LHC energies 
\cite{ref:alicedata}. 
%
\begin{figure}[th!]
\vspace*{-0.5cm}
\begin{center}
\epsfig{figure=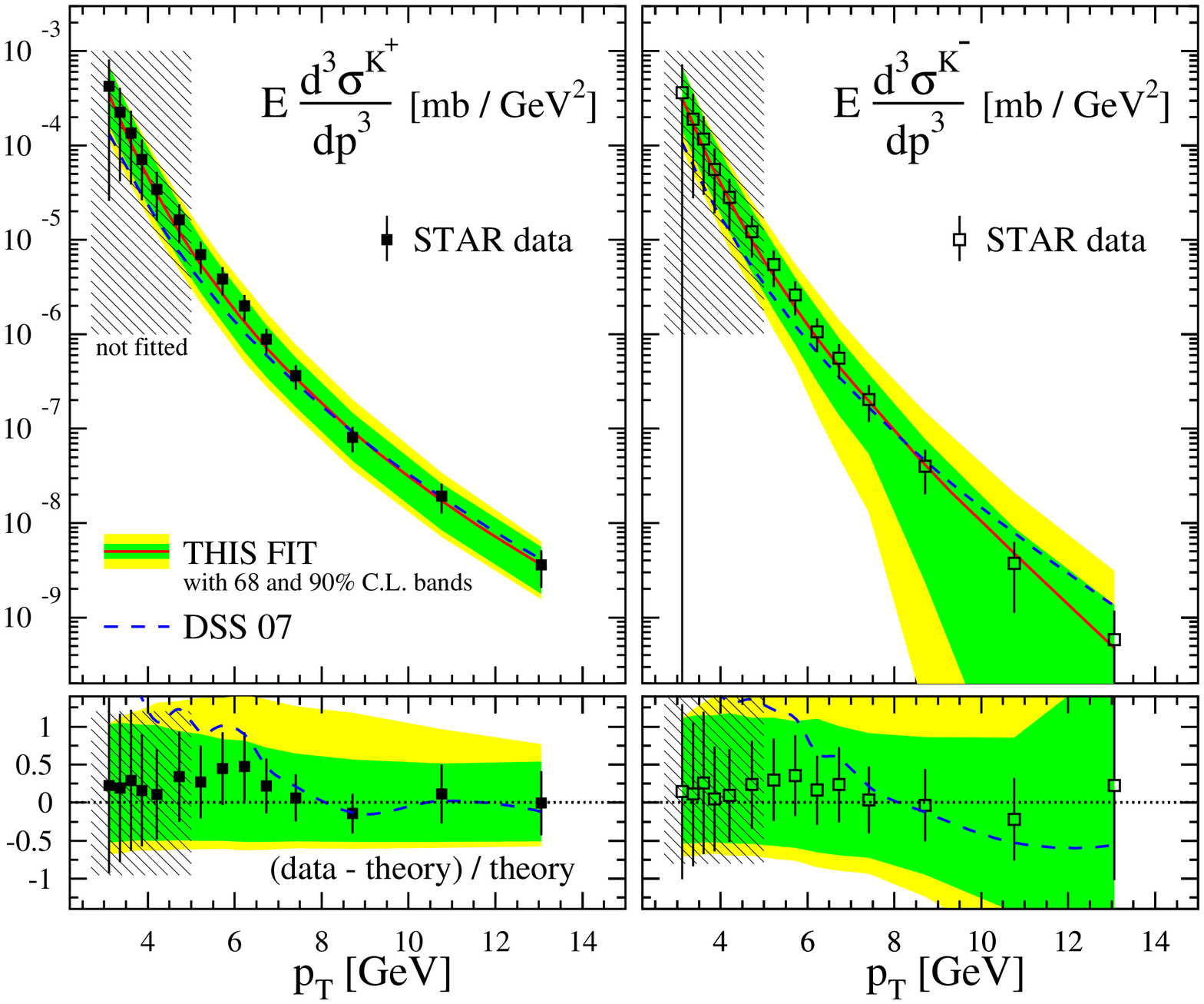,width=0.5\textwidth}
\end{center}
\vspace*{-0.5cm}
\caption{Comparison of our NLO results for the $K^+$ (left-hand-side) 
and $K^-$ (right-hand-side) cross sections in $pp$ collisions at 
$\sqrt{S}=200\,\mathrm{GeV}$ and mid rapidity with the {\sc{Star}} data \cite{ref:starratio11}.
The inner and outer shaded bands correspond to uncertainty estimates at $68\%$ and $90\%$ C.L.,
respectively. Also shown are the results obtained with the DSS~07 set of kaon FFs (dashed lines).
The lower panels show the corresponding results for (data-theory)/theory.
\label{fig:pp-ka-star}}
\vspace*{-0.75cm}
\begin{center}
\epsfig{figure=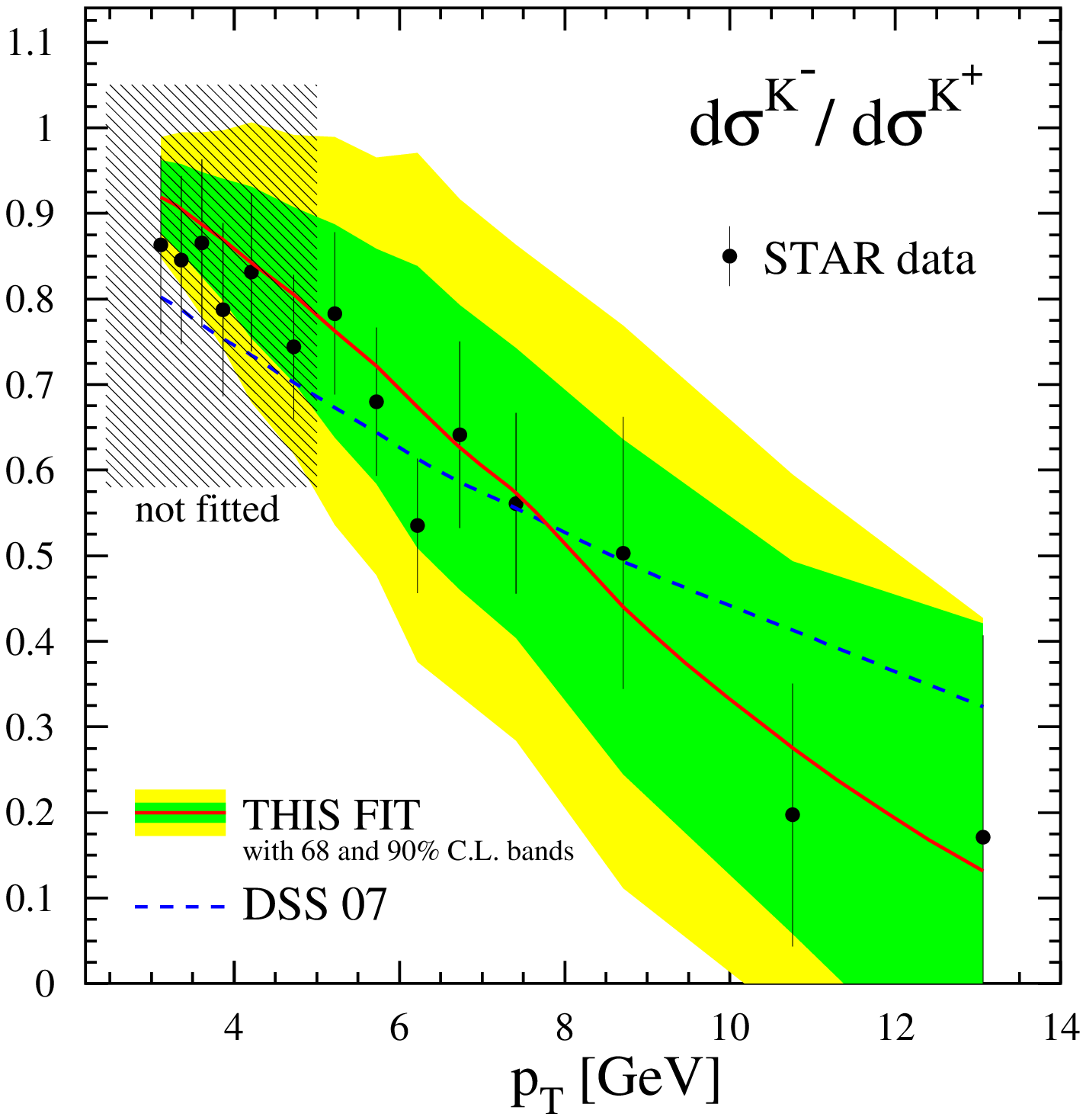,width=0.5\textwidth}
\end{center}
\vspace*{-0.5cm}
\caption{Same as in Fig.~\ref{fig:pp-ka-star} but now for the 
$K^-/K^+$ cross section ratio.
\label{fig:pp-ka-ratio-star}}
\end{figure} 

Due to the complexity of the underlying hard-scattering processes at NLO accuracy 
\cite{ref:pp-nlo}, the repeated numerical evaluation of single-inclusive hadron production
yields in $pp$ collisions in a $\chi^2$-minimization procedure is very time-consuming. 
The use of a fast, grid-based method 
to implement the relevant NLO expressions efficiently and without the 
need of any approximations such as ``K-factors'' is indispensable here.
As in all our various previous global analyses \cite{ref:dss,ref:dss2,ref:eta,ref:dssv}, 
and for the implementation of the SIDIS multiplicities in Sec.~\ref{sec:sidis-data},
we resort to the well-tested method based on Mellin moments, see Ref.~\cite{ref:mellin2}.

Data for inclusive particle spectra at not too large values of $p_T$
in $pp$ collisions draw their relevance in a global fit from the dominance of
gluon-induced processes. Many of the observed hadrons stem from the hadronization
of gluons both at RHIC and LHC energies \cite{ref:us-lhc}. Hence, such data are
expected to provide invaluable information on the otherwise (i.e., by SIA and SIDIS data)
only weakly constrained gluon FF $D_g^{K^{+}}$. 

In our corresponding global analysis of parton-to-pion FFs in Ref.~\cite{deFlorian:2014xna}
we have found some tension between the $p_T$ spectra of neutral pions
measured at $\sqrt{S}=200\,\mathrm{GeV}$ at RHIC 
and results from the LHC at much higher c.m.s.\ energies up to $\sqrt{S}=7\,\mathrm{TeV}$.
In some sense this was already anticipated by comparisons of LHC data to expectations 
computed with the previous DSS~07 sets of FFs \cite{ref:dss,ref:dss2}, 
which are known to describe the  RHIC data nicely down to, perhaps unexpectedly small 
$p_T\simeq 1.5\,\mathrm{GeV}$ \cite{ref:dss} but were found to grossly overshoot yields for
both neutral pions and unidentified charged hadrons (that are dominated by pions)
at essentially all $p_T$ values \cite{ref:helenius}.
In particular, at smallish $p_T$ values, below about $5\,\mathrm{GeV}$, the 
data from RHIC and the LHC appear to be mutually exclusive in a global QCD analysis.
Since the origin of this discrepancy could not be traced and we did not want to remove
arbitrarily either of the data sets from the analysis, a cut $p_T \ge 5\,\mathrm{GeV}$ 
was introduced in our fit to remedy the tension, see Ref.~\cite{deFlorian:2014xna}.
Since we wish to analyze data for the kaon-to-pion ratio, utilizing the latest DSS~14 pion FFs,
we decided to proceed with the same cut on $p_T$ in the present analysis of kaon FFs.

Figure~\ref{fig:pp-ka-star} shows the data from the {\sc Star} Collaboration \cite{ref:starratio11}
for single-inclusive charged kaon yields at mid rapidity compared to the results of our fit
at NLO accuracy. In Fig.~\ref{fig:pp-ka-ratio-star} the corresponding cross section ratio
is displayed. Since theoretical scale and PDF ambiguities partially cancel in the $K^-/K^+$ ratio,
we decided to use it our fit along with the cross section data for $K^+$ in the left panel
of Fig.~\ref{fig:pp-ka-star}. To avoid double counting of the same data in the fit, we discard
the $K^-$ cross section in the $\chi^2$-minimization but illustrate how well the data are
described in the right-hand-side of Fig.~\ref{fig:pp-ka-star}.
As can be seen from both figures and Tab.~\ref{tab:exppiontab}, the quality of the fit
is very good, even when extrapolated to the $p_T$-region below $5\,\mathrm{GeV}$.
The latter feature indicates that unlike for pions \cite{deFlorian:2014xna}
there is considerably less tension with the LHC data from the 
{\sc Alice} Collaboration; see below.
Calculations based on the old DSS~07 set of FFs provide a fair description of the
{\sc Star} charged kaon data but the $p_T$ slope is somewhat off.

\begin{figure}[ht!]
\vspace*{-0.75cm}
\begin{center}
\epsfig{figure=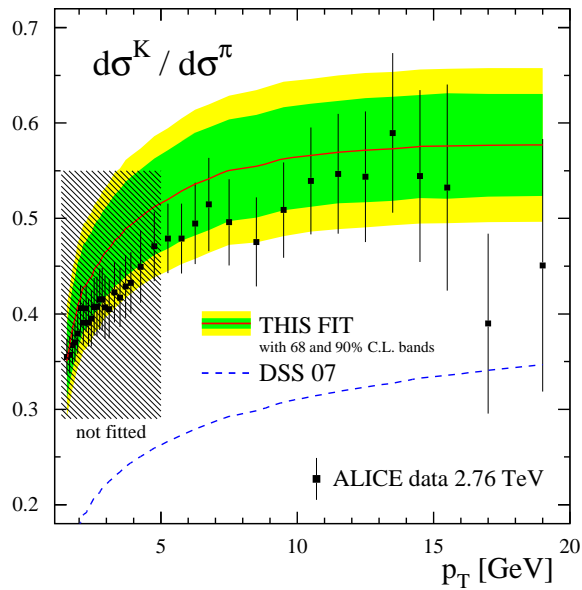,width=0.5\textwidth}
\end{center}
\vspace*{-0.5cm}
\caption{Ratio of the charged kaon to charged pion cross section at
$\sqrt{S}=2.76\,\mathrm{TeV}$ as measured by {\sc Alice} \cite{ref:alicedata}
compared to our NLO results (solid line). The pion cross section is computed
with the DSS~14 set \cite{deFlorian:2014xna}. The dashed line illustrates the result obtained with the 
old DSS~07 sets of FFs for both pions and kaons.
The inner and outer shaded bands correspond to uncertainty estimates 
at $68\%$ and $90\%$ C.L., respectively. 
\label{fig:alice}}
\end{figure} 
Finally, in Fig.~\ref{fig:alice} we show the charged kaon to charged pion cross section ratio 
as a function of the transverse momentum $p_T$ as measured by the {\sc Alice} Collaboration
in $pp$ collisions at mid rapidity at a c.m.s.\ energy $\sqrt{S}$ of $2.76\,\mathrm{TeV}$ \cite{ref:alicedata}. 
The ratio is estimated by dividing the cross section computed with the parton-to-kaon fragmentation
functions obtained in the present analysis by the one 
obtained with the DSS~14 set of parton-to-pion FFs of Ref.~\cite{deFlorian:2014xna},
including the quoted normalization shift for the {\sc Alice} pion data. 
As can be seen, the current description of the data is much better than the one achieved by
the previous DSS~07 sets of pion and kaon FFs (dashed line) which turns out to be way to small
in the entire range of $p_T$. One reason is the much reduced gluon-to-pion FF in 
the DSS~14 set \cite{deFlorian:2014xna} as compared to DSS~07, which pushes the kaon-to-pion
ratio up. In addition, the new fit has a larger gluon-to-kaon FF than in our previous
DSS~07 analysis as can be inferred from Fig.~\ref{fig:ff-at-10}.
 
The inner and outer shaded bands in Figs.~\ref{fig:pp-ka-star} - \ref{fig:alice}
represent our uncertainty estimates at $68\%$ and $90\%$ C.L., respectively.
The bands are considerably wider than for the corresponding kinematics for pion yields,
see Figs.~9-11 in Ref.~\cite{deFlorian:2014xna}.
In addition, there are theoretical uncertainties from the choice of the factorization and
renormalization scales and the set of PDFs in the cross section calculations. For the
results shown in the figures, we use a common scale $\mu_f=\mu_r=p_T$ and, 
as for SIDIS multiplicities, the MMHT set of PDF \cite{Harland-Lang:2014zoa}. 
Since the relevant kinematics and the dominance of gluons is very similar 
to the case of single-inclusive pion production at RHIC and the LHC, also the scale and
PDF uncertainties for kaons are similar, see Figs.~9-11 in Ref.~\cite{deFlorian:2014xna}
for estimates. For kaons, however, the uncertainty estimates at $68\%$ and $90\%$ C.L.\
shown in Figs.~\ref{fig:pp-ka-star} - \ref{fig:alice} are now the dominant ones, which
basically reflects the fact that the experimental data for kaon production
is less accurate that those for pions.

\section{Summary and Outlook}
%
We have presented a new, comprehensive global QCD analysis of parton-to-kaon fragmentation 
functions at next-to-leading order accuracy including the latest experimental information.
The analyzed data comprise single-inclusive kaon production in semi-inclusive electron-positron
annihilation, deep-inelastic scattering, and proton-proton collisions and span
energy scales ranging from about $1\,\mathrm{GeV}$ up to the mass of the $Z$ boson.
The very satisfactory and simultaneous description of all data sets within the estimated
uncertainties strongly supports the validity of the underlying theoretical framework based on pQCD and, in 
particular, the notion of factorization and universality for parton-to-kaon fragmentation 
functions.

Compared to our previous analysis of kaon fragmentation functions in 2007, which was based on much less precise 
and copious experimental inputs, and to which we have made extensive comparisons throughout this work,
we now obtained a significantly better fit, as measured in terms its the global $\chi^2$, 
using the same functional form as before with only a few additional fit parameters. 
While most of the favored and unfavored quark-to-kaon fragmentation functions are by and large
similar to our previous results, perhaps the most noteworthy change is a larger gluon-to-kaon
fragmentation function, which can be tested and constrained further by upcoming data from the LHC experiments.

The wealth of new data included in our updated global analysis
allow for the first time to perform a reliable estimate of uncertainties for
parton-to-kaon fragmentation functions based on the standard iterative Hessian method.
The availability  of Hessian sets will significantly facilitate the propagation of these uncertainties to
other observables with identified kaons.
The obtained uncertainties are still sizable 
in the kinematic regions covered and constrained by data and they quickly deteriorate beyond.
They range at best from about twenty to thirty percent for the total strange quark fragmentation function 
and from ten to twenty five percent for the total $u$ quark and the gluon fragmentation functions.
Another new asset of the current analysis was to analytically determine the optimum normalization
shift for each data set in the fit, which greatly facilitated the fitting procedure.

The newly obtained kaon fragmentation functions and their uncertainty estimates will be a crucial 
ingredient in future global analyses of both helicity and transverse-momentum dependent parton
densities, which heavily draw on data with identified kaons in the final-state.
Our results will also serve as the baseline in heavy-ion and proton-heavy ion collisions,
where one of the main objectives is to quantify and understand possible modifications of
hadron production yields by the nuclear medium.
Since pions and kaons constitute by far the largest fraction in frequently measured 
yields of {\em unidentified} charged hadrons, our newly updated sets of fragmentation functions
for both will be a good starting point for a future global QCD analysis of
fragmentation functions for unidentified hadrons. It will be interesting to quantify how much
room is left for other hadron species, in particular, for protons.
  
Further improvements of parton-to-kaon fragmentation functions from the theory side
should include an improved treatment of heavy quark-to-kaon fragmentation functions,
likely along similar lines as for heavy flavor parton densities. Also, the impact
of higher order corrections beyond the next-to-leading order accuracy and all-order 
resummations should be explored. Some results in all these directions have become available
recently, however, complete next-to-next-to-leading order corrections are currently
only available for an analysis of electron-positron annihilation data.
Also the potential bias from the choice of parton distribution functions in the
extraction of fragmentation functions from, in particular, semi-inclusive deep-inelastic scattering
at not too large scales is worth investigating further. Ultimately, a combined global analysis
of parton distribution and fragmentation functions must be the goal.

On the experimental side, the LHC will continue to provide new, valuable data on 
identified and unidentified hadron spectra.
To improve our knowledge on the flavor separation of fragmentation functions, it will be 
of paramount importance to fully utilize the unprecedented 
capabilities in semi-inclusive deep-inelastic scattering that will open up
at a future electron-ion collider, a project that is currently under scrutiny in the U.S. 

\section*{Acknowledgments}
%
We are grateful to G. Schnell and E. C. Aschenauer ({\sc Hermes}), 
M.\ Leitgab and R.\ Seidl ({\sc Belle}) and F.\ Kunne, E. Seder 
({\sc Compass}) for helpful discussions about their measurements.
We warmly acknowledge vivid discussions with E.\ Leader about the
consistency of SIDIS data.
The work of RJHP is partially supported by CONACyT and PROFAPI 2015
grant No.\ 121.
This work was supported in part by CONICET, ANPCyT,
and the Institutional Strategy of the University of T\"{u}bingen 
(DFG, ZUK 63).  


%
\end{document}